\documentclass[sigconf]{acmart}

\copyrightyear{2021} 
\acmYear{2021} 
\setcopyright{none}\acmConference[ESEM '21]{ACM/IEEE International Symposium on Empirical Software Engineering and Measurement (ESEM)}{October 11--15, 2021}{Bari, Italy}
\acmBooktitle{ACM / IEEE International Symposium on Empirical Software Engineering and Measurement (ESEM) (ESEM '21), October 11--15, 2021, Bari, Italy}
\acmDOI{10.1145/3475716.3475780}
\acmISBN{978-1-4503-8665-4/21/10}

\usepackage{xspace}
\usepackage{ifthen}
\usepackage[normalem]{ulem} 
\usepackage{array, booktabs}
\usepackage{ifthen}
\usepackage[most]{tcolorbox}
\usepackage{paralist} 
\usepackage{fontawesome}      

\newcolumntype{R}[2]{
    >{\adjustbox{angle=#1,lap=\width-(#2),margin*=0.4em 0em 0em 0em}\bgroup}
    l
    <{\egroup}
}

\let\oldFootnote\footnote
\newcommand\nextToken\relax
\renewcommand\footnote[1]{%
    \oldFootnote{#1}\futurelet\nextToken\isFootnote}
\newcommand\isFootnote{%
    \ifx\footnote\nextToken\textsuperscript{,}\fi}

\newcommand{\id}[1]{$-$Id: scgPaper.tex 32478 2010-04-29 09:11:32Z oscar $-$}

\newcommand{\ie}{\emph{i.e.},\xspace}
\newcommand{\eg}{\emph{e.g.},\xspace}
\newcommand{\etal}{\emph{et al.}\xspace}

\newcommand{\smellText}[1]{\emph{#1}}
\newcommand{\Issue}{\emph{Issue:}\xspace}

\newcommand{\Symptoms}{\emph{Symptom:}\xspace}

\newcommand{\code}[1]{\texttt{#1}}

\newcolumntype{L}[1]{>{\raggedright\arraybackslash}m{#1}} 

\def\BibTeX{{\rm B\kern-.05em{\sc i\kern-.025em b}\kern-.08em
	T\kern-.1667em\lower.7ex\hbox{E}\kern-.125emX}}
	
\newboolean{showedits}
\setboolean{showedits}{true} 
\ifthenelse{\boolean{showedits}}
{
	\newcommand{\del}[1]{\textcolor{red}{\sout{#1}}} 
	\newcommand{\nbe}[3]{
		{\colorbox{#3}{\bfseries\sffamily\scriptsize\textcolor{white}{#1}}}
		{\textcolor{#3}{\sf\small$\blacktriangleright$\textit{#2}$\blacktriangleleft$}}}
}{
	\newcommand{\del}[1]{} 
	
	\newcommand{\nbe}[3]{}
}

\ifthenelse{\boolean{showedits}}
{
	\newtcolorbox{inserted}{
		title=Inserted text:,
		colframe=blue,colback=blue!5!white,
		breakable,
		leftrule=0mm, 
		bottomrule=0mm,
		rightrule=0mm,
		toprule=0mm,
		arc=0mm, outer arc=0mm,
		oversize
	}
	\newtcolorbox{deleted}{
		title=Deleted text:,
		colframe=red,colback=red!5!white,
		breakable,
		leftrule=0mm, 
		bottomrule=0mm,
		rightrule=0mm,
		toprule=0mm,
		arc=0mm, outer arc=0mm,
		oversize
	}
	\newtcolorbox{refactored}{
		title=Rewritten text:,
		colframe=blue,colback=red!5!white,
		breakable,
		leftrule=0mm, 
		bottomrule=0mm,
		rightrule=0mm,
		toprule=0mm,
		arc=0mm, outer arc=0mm,
		oversize
	}
}{

}

\newboolean{showcomments}
\setboolean{showcomments}{true}
\ifthenelse{\boolean{showcomments}}
{\newcommand{\nbc}[3]{
		{\colorbox{#3}{\bfseries\sffamily\scriptsize\textcolor{white}{#1}}}
		{\textcolor{#3}{\sf\small$\blacktriangleright$\textit{#2}$\blacktriangleleft$}}}
	}
{\newcommand{\nbc}[3]{}
	}

\definecolor{source}{gray}{0.9}
\newcommand\keyword[1]{[#1]}

\newcommand{\newevenside}{
	\ifthenelse{\isodd{\thepage}}{\newpage}{
		\newpage
		\phantom{placeholder} 
		\thispagestyle{empty} 
		\newpage
	}
}

\DeclareGraphicsExtensions{.pdf,.png}
\graphicspath{{images/}}

\newboolean{isblinded}
\setboolean{isblinded}{false}
\ifthenelse{\boolean{isblinded}}
{\newcommand\blind[1]{BLINDED\xspace}}
{\newcommand\blind[1]{#1\xspace}}

\begin{document}

\hyphenation{op-tical net-works semi-conduc-tor}

\title{Security Smells Pervade Mobile App Servers}

\author{\blind{Pascal Gadient}}
\author{\blind{Marc-Andrea Tarnutzer}}
\author{\blind{Oscar Nierstrasz}}
\affiliation{%
  \institution{\blind{Software Composition Group, University of Bern}}
  \city{\blind{Bern}}
  \country{\blind{Switzerland}}
  }

\author{\blind{Mohammad Ghafari}}
\affiliation{%
\institution{\blind{University of Auckland}}
\city{\blind{Auckland}}
\country{\blind{New Zealand}}}
\email{m.ghafari@auckland.ac.nz}

\begin{abstract}
\keyword{Background}
Web communication is universal in cyberspace, and security risks in this domain are devastating.
\keyword{Aims}
We analyzed the prevalence of six security smells in mobile app servers, and we investigated the consequence of these smells from a security perspective.
\keyword{Method}
We used an existing dataset that includes 9\,714 distinct URLs used in 3\,376 Android mobile apps.
We exercised these URLs twice within 14 months and investigated the HTTP headers and bodies.
\keyword{Results}
We found that more than 69\% of tested apps suffer from three kinds of security smells, and that unprotected communication and misconfigurations are very common in servers.
Moreover, source-code and version leaks, or the lack of update policies expose app servers to security risks.
\keyword{Conclusions}
Poor app server maintenance greatly hampers security.
\end{abstract}



\maketitle

\section{Introduction}
\label{sec:introduction}
Globally accessible, reliable, and scalable web services are on the rise, with more than 24\,000 currently known public web APIs.\footnote{\url{https://www.programmableweb.com/apis/directory}}
Likewise, native apps are starting to decline while web apps arise that depend on application servers.\footnote{\url{https://www.forbes.com/sites/victoriacollins/2019/04/05/why-you-dont-need-to-make-an-app-a-guide-for-startups-who-want-to-make-an-app/}}
Additionally during the past years, the complexity of developing a web-enabled app has massively increased due to the growing number of involved application frameworks, programming languages, and supported device categories, \eg desktops, notebooks, tablets, smartphones, and wearables.

Web communication has already received much attention in the security community, leading to improved tool support.
For example, programs exist that can continuously monitor web APIs to ensure that an app remains compliant with the API specification~\cite{Wittern:2017}, and tool support to mitigate insecure communication channels has been built into recent releases of the Android ecosystem.\footnote{\url{https://developer.android.com/training/articles/security-config}}
Generally speaking, the existing literature has covered the transmitted payload by using client-side static~\cite{Gordon:2015} and dynamic analysis techniques~\cite{Rapoport:2017}, server side analyses of web service source code~\cite{Mendoza:2018a} as well as connection properties, \eg the URL~\cite{Zuo:2017} or two-factor authentication~\cite{Tang:2015}.
Another topic that has received extensive attention is that of hard-coded credentials in apps that may allow adversaries unrestricted access to the relevant infrastructure~\cite{Zhou:2015,Rahman:2019}.

Unfortunately, the server configurations of off-the-shelf application servers have received much less attention.
A recent work has identified eight web API security smells, but did not assess their prevalence~\cite{Gadient:2020}.
These smells are \emph{symptoms in the code that signal the prospect of a security vulnerability}~\cite{Ghafari:2017}.
In this work we assess app servers that are used for communicating with mobile apps.
We investigated the presence of six app server security smells, and the corresponding server maintenance activity based on the dataset that contains 9\,714 distinct URLs that were used in 3\,376 apps.
We address the following research questions:
\newcommand{\rqthree}{\emph{What is the prevalence of the server side security smells in the web communication of mobile apps?}\xspace}
\newcommand{\rqsix}{\emph{What is the relationship between security smells and app server maintenance?}\xspace}

\textbf{RQ$_{1}$}: \rqthree
We found 231 URLs from 44 apps that leak the source code of the web service implementation if processing errors occur.
We can further confirm that most app servers communicate with apps over insecure HTTP connections~\cite{Possemato:2020}, and fail to enforce use of the HTTP strict transport security policy.
Finally, we found that on average almost every second app server suffers from version information leaks.

\textbf{RQ$_{2}$}: \rqsix 
In particular, we are interested in configuration changes, because they provide insights into established maintenance processes of mobile app servers.
Based on the collected HTTP header information from two measurements over 14 months, we evaluated what software changes are introduced by system administrators.
We observed that servers are usually set up once and never touched again, yielding severe security risks.
For instance, criminals can attack outdated app servers by exploiting vulnerabilities listed in public databases or illicit websites.
On the positive side, we noted that version upgrades are much more common than version downgrades, and that developers occasionally use Cloudflare to protect their infrastructure against adversaries, especially for non-JSON-based app servers.

In summary, this work reveals the prevalence of insecure app server configurations accessed by Android mobile apps, and their maintenance protocol.
The list of apps that we analyzed in this study is available online,\footnote{\url{https://doi.org/10.6084/m9.figshare.14981061}} and we share the aggregated data for research purposes on request due to the contained sensitive information such as credentials, API keys, and email addresses.

The remainder of this paper is organized as follows.
We report on app server security smells relevant for this work in Section~\ref{sec:web-api-security-smells}.
We describe the dataset used for our app server investigations in Subsection~\ref{sec:methodology}.
We provide the prevalence of app server security smells in Subsection~\ref{sec:prevalence-security-smells}, and we shed light on server infrastructure maintenance in Subsection~\ref{sec:maintenance-of-server-infrastructure}.
Finally, we recap the threats to validity in Section~\ref{sec:threats-to-validity}, and we summarize related work in Section~\ref{sec:related-work}.
We conclude this paper in Section~\ref{sec:conclusion}.

\section{Security Smells in App Servers}
\label{sec:web-api-security-smells}
In this section, we briefly explain six of the eight security smells that we identified in previous work~\cite{Gadient:2020}.
The two remaining security smells, \ie \emph{Credential leak} and \emph{Embedded languages} are not within the scope of this study, because they require a deep understanding of the app and the context where they occur.

\subsection{Insecure transport channel}
Web communication relies on HTTP or HTTPS; both variants exist in app server configurations.
\Issue 
HTTP does not provide any security; neither the address, nor the header information or the payload are encrypted.
\Symptoms 
The URL begins with \texttt{http://}.

\subsection{Disclosure of source code}
Error messages leak valuable information regarding the implementation of a running system.
\Issue 
Error messages that include the relevant stack trace are transmitted as plain text in the server's message response body.
Such a message reveals information like the used method names, line numbers, and file paths disclosing the internal file system structure and configuration of the server.
\Symptoms 
The returned HTTP body contains a stack trace or a code snippet that shows the problematic code.
The structure of such data depends on the used framework, however terms related to application crashes are common, \eg 
``stack,'' ``trace,'' and ``error.''

\subsection{Disclosure of version information}
Besides useful connection parameters, HTTP headers leak information regarding the software architecture and configuration of a running system.
\Issue 
Outdated software suffers from severe security vulnerabilities.
For instance, a server that returns \code{X-Powered\-By: PHP/5.5.23} in the response header uses a PHP version that is at the time of writing more than 6 years old, and a quick search in the Common Vulnerabilities and Exposures (CVE) database shows that this framework suffers from 65 known security vulnerabilities, six of which received the most severe impact score of 10.\footnote{\url{https://www.cvedetails.com/vulnerability-list/vendor_id-74/product_id-128/version_id-183021/PHP-PHP-5.5.23.html}}
\Symptoms 
One of the following keys exists in the response header: \code{engine}, \code{server}, \code{x-aspnet-version}, or \code{x-powered-by}.

\subsection{Lack of access control}
Authentication by a user name and a password provides tailored experiences to end users, \eg individual chat logs or friend lists, and at the same time enables access control to separate and protect sensitive user data.
\Issue 
The access to sensitive data or actions is not restricted by a sane authentication mechanism such as a user name and password pair, instead, easy-to-forge identifiers or no identification data at all are used to secure the access.
\Symptoms 
A server does not respond with the status code \emph{401 Unauthorized} or \emph{403 Forbidden}.
In other words, the server successfully responds without asking for any credentials.

\subsection{Missing HTTPS redirects}
We found servers that do not redirect clients to encrypted connections although they would have been supported.
\Issue 
App servers do not redirect incoming HTTP connections to HTTPS when legacy apps try to connect.
\Symptoms 
For an HTTP request, a server does not deliver an HTTP \code{3xx} redirect message which points to the corresponding HTTPS implementation of the web application.
For an HTTPS request, a server delivers a HTTP redirect message.

\subsection{Missing HSTS}
HTTP header information is used to properly set up the connection by specifying various communication parameters, \eg the acceptable languages, the used compression, or the enforcement of HTTPS for future connection attempts, a feature which is called \emph{HTTP Strict Transport Security (HSTS)}.
HSTS provides protection against HTTPS to HTTP downgrading attacks, \ie when a user once accessed a web resource in a secure environment (at home or work), the client knows that the resource needs to be accessed \emph{only} through HTTPS.
If this is not possible, \eg at an airport at which an attacker tries to perform MITM attacks, the client will display a connection error.
Hence, HSTS should be used in combination with HTTP to HTTPS redirects, because the HSTS header is only considered to be valid when sent over HTTPS connections.
\Issue 
Servers either do not leverage the HSTS feature, or they do not use the recommended parameters.
\Symptoms 
A server does not deliver the HTTP HSTS header \code{Strict-Trans\-port-Security: max-age=31536000; includeSubDomains} for an HTTPS request.

\section{Empirical Study}
For this empirical study we evaluated all URLs from the dataset according to the security smell symptoms described in the previous section.
We collected the data twice:
the initial download of HTTP headers and bodies was performed in June-2019 whereas additional data, \ie the authorization errors and up-to-dateness, was retrieved in August-2020.
The duration of 14 months is arbitrary but long enough to ensure developers have to update their software infrastructure.

\subsection{Dataset}
\label{sec:methodology}
We build on our previous work and dataset~\cite{Gadient:2020} in which we manually inspected Android apps to identify which APIs developers use to call web services, and how they are used.
We then took advantage of this information to develop a tool to automatically extract and reconstruct string variables and the assigned values, the server URLs and their corresponding HTTP request headers statically from the apps.
Using this information, we analyzed the reconstructed app server data and tried to establish connections to the corresponding servers from which they gathered additional information for analysis, \ie from HTTP response headers.

The apps from the dataset are randomly collected from those that use Android's internet permission.
For closed-source apps we mined the free apps on the \emph{Google Play} store, and for the open-source apps we relied on the \emph{F-Droid} software repository.\footnote{\url{https://f-droid.org}}
For each app, we removed the duplicates, \ie apps with the same package identifier, but different version numbers, and kept only the most recent version of the app.
We also included the partial results of the apps whose analysis was incomplete and could not finish in time, ultimately resulting in an analysis result for 303 open-source, and 3\,073 closed-source apps in the dataset.

The apps in the dataset come from 48 different Google Play store categories.
Most of them belong to \code{EDUCATION} (317 apps) and \code{TOOLS} (292 apps), however, a majority (574) have a \code{GAMES}-related tag.
Interestingly, work-related apps are common in the dataset (335 apps).
The top five categories whose apps contain the largest number of distinct URLs are \code{EDUCATION} (1\,555 URLs), \code{LIFESTYLE} (1\,027 URLs), \code{BUSINESS} (995 URLs), \code{ENTERTAINMENT} (704 URLs), and \code{PRODUCTIVITY} (619 URLs).
\begin{figure}
\centering
\includegraphics[scale = 0.39, trim = 2cm 15.5cm 2cm 1.25cm]{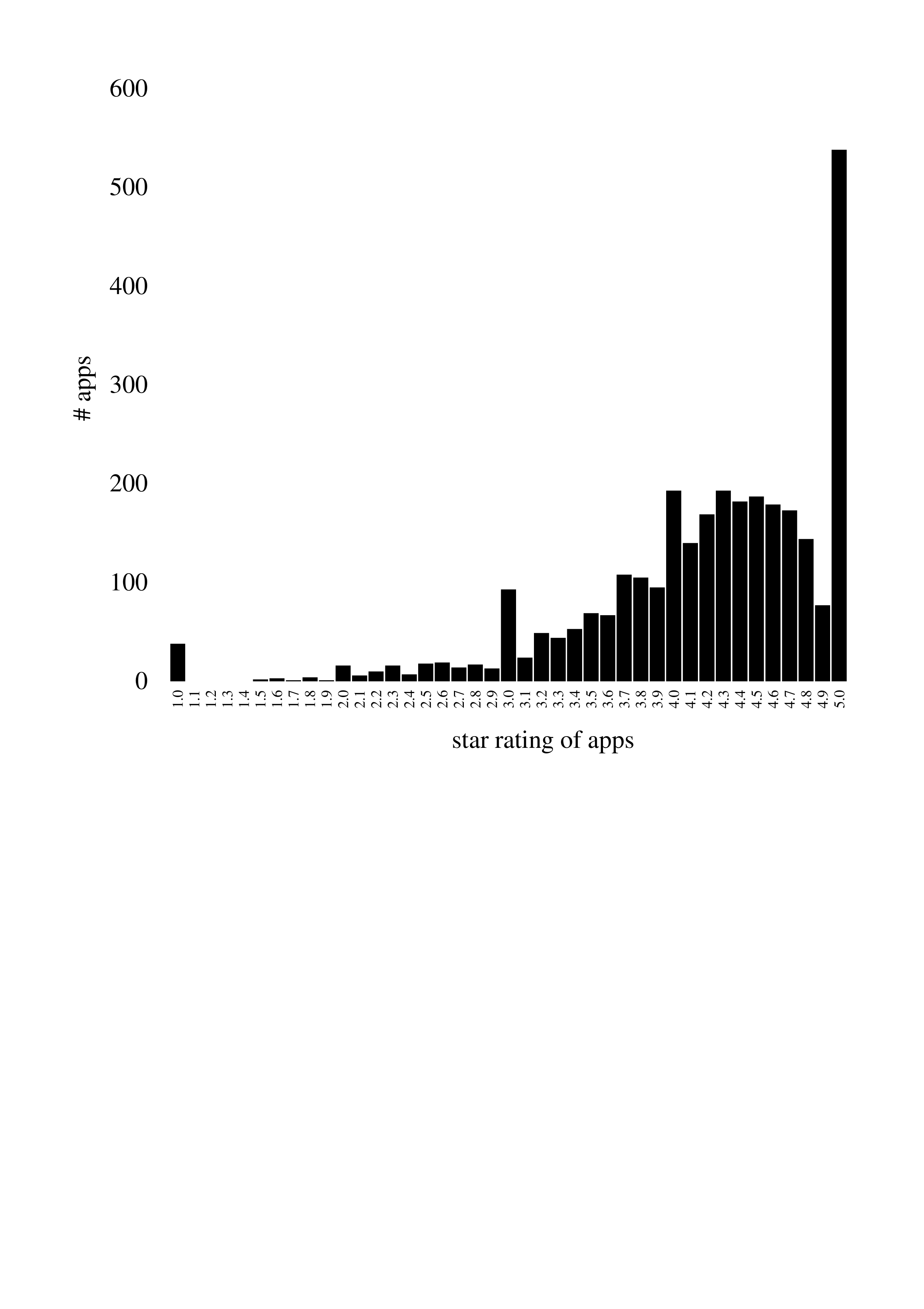}
\caption{Star ratings for the Google Play apps in the dataset}
\label{fig:p_playstore_starrating}
\end{figure}
\begin{figure}
\centering
\includegraphics[scale = 0.39, trim = 2cm 15.5cm 2cm 1.25cm]{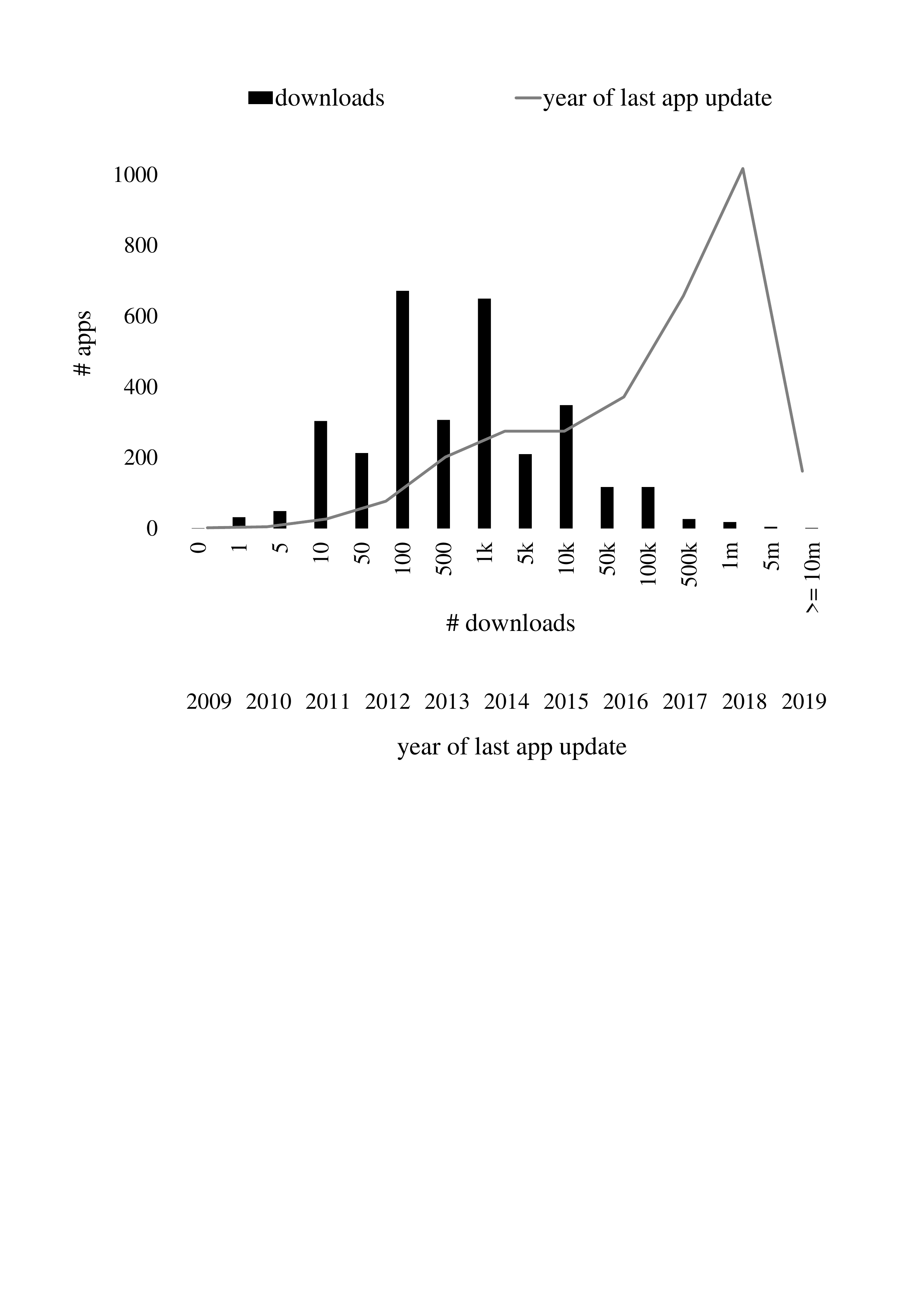}
\caption{The popularity and developer support for the Google Play apps in the dataset}
\label{fig:p_playstore_data}
\end{figure}
As shown in~\autoref{fig:p_playstore_starrating}, almost 94\% of the apps received a star rating of 3.0 or higher.
Surprisingly, apps with a five star rating are more prevalent than apps in any other category.
The apps have an average star rating of 4.2 stars and a median rating of 4.3 stars.
\autoref{fig:p_playstore_data} presents the number of app downloads and the timeliness of app updates.
The y-axis denotes the number of apps in each category.
In contrast, the primary x-axis with the bars indicates the app downloads, and the secondary x-axis with the line indicates the time of the last app update.
We can see that most apps achieved between 100 and 1\,000 downloads, and barely any app was downloaded more than 1 million times.
Regarding the app updates, most of the apps received an update in 2018.
Therefore, we see that most vendors update their apps only a few times a year, because we collected the statistics separately in 2019.

We then exercised every URL in the dataset and collected the HTTP header and body of each server response.
Eventually, we processed 1\,230 open-source URLs and 8\,486 closed-source URLs.
We realized that many app servers do not leverage JSON, but instead they use, for example, XML or plain HTTP communication.
Because we were interested whether there exist any differences for data-centric app servers, we split the results into four different groups.
We report our findings based on closed-source and open-source apps, and we also separate between JSON and non-JSON app servers.
We favored the JSON data format, because it was much more commonly used for communication than the others.
Therefore, we partitioned the open-source URLs into 1\,171 non-JSON URLs and 59 JSON URLs.
Accordingly, we partitioned the closed-source URLs into 7\,997 non-JSON URLs and 489 JSON URLs.

We were particularly interested in information such as operating system identifiers, used software modules, and version numbers.
Hence, we crafted a number of search queries to detect occurrences of such features.
The relevant features, \ie security smells, and the results are part of the discussion in the subsequent subsections.

\begin{figure}
\centering
\includegraphics[scale = 0.39, trim = 2cm 15.5cm 2cm 1.25cm]{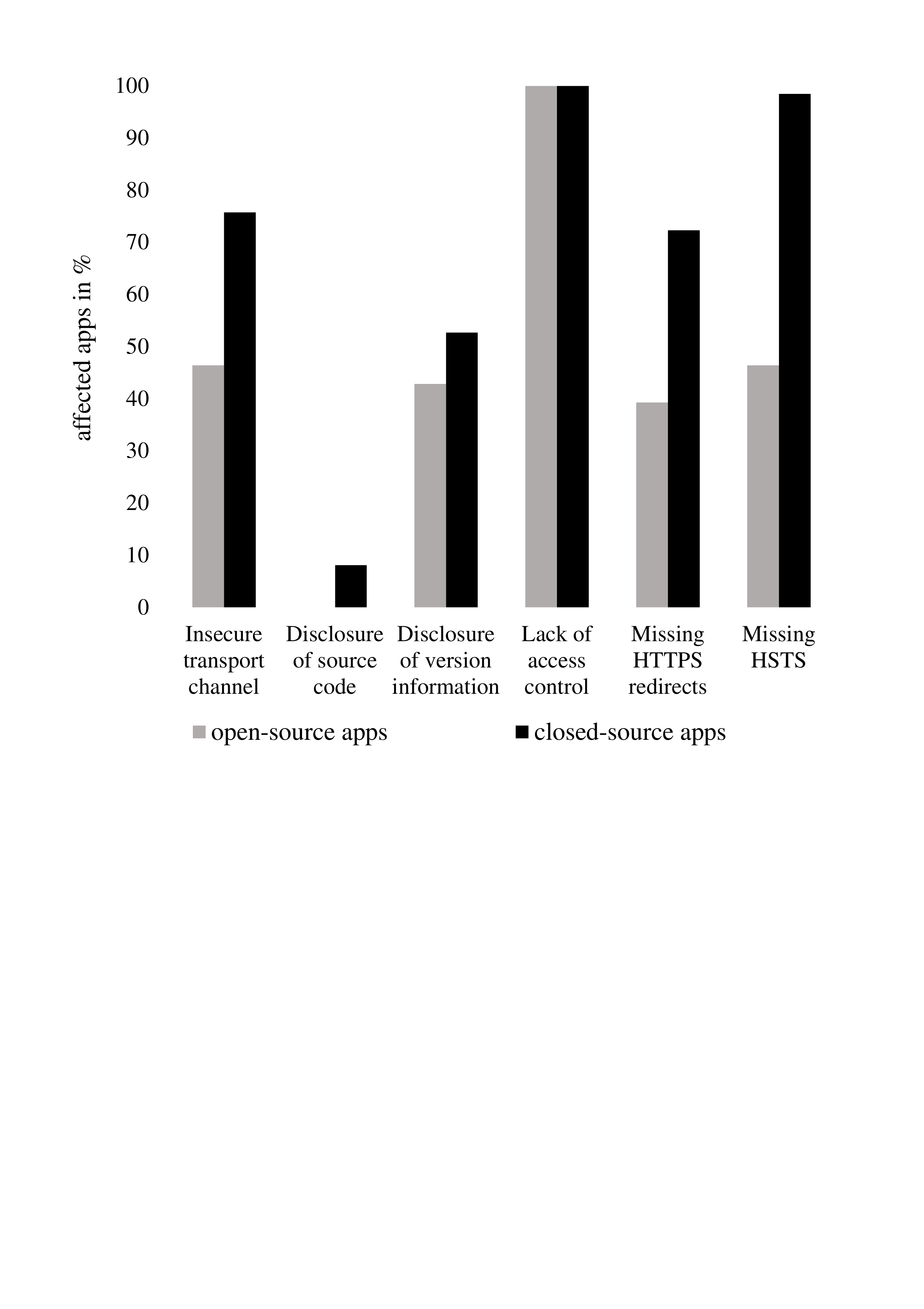}
\caption{Prevalence of app server smells in apps considering JSON communication}
\label{fig:p_smells_json}
\end{figure}

\begin{figure}
\centering
\includegraphics[scale = 0.39, trim = 2cm 15.5cm 2cm 1.25cm]{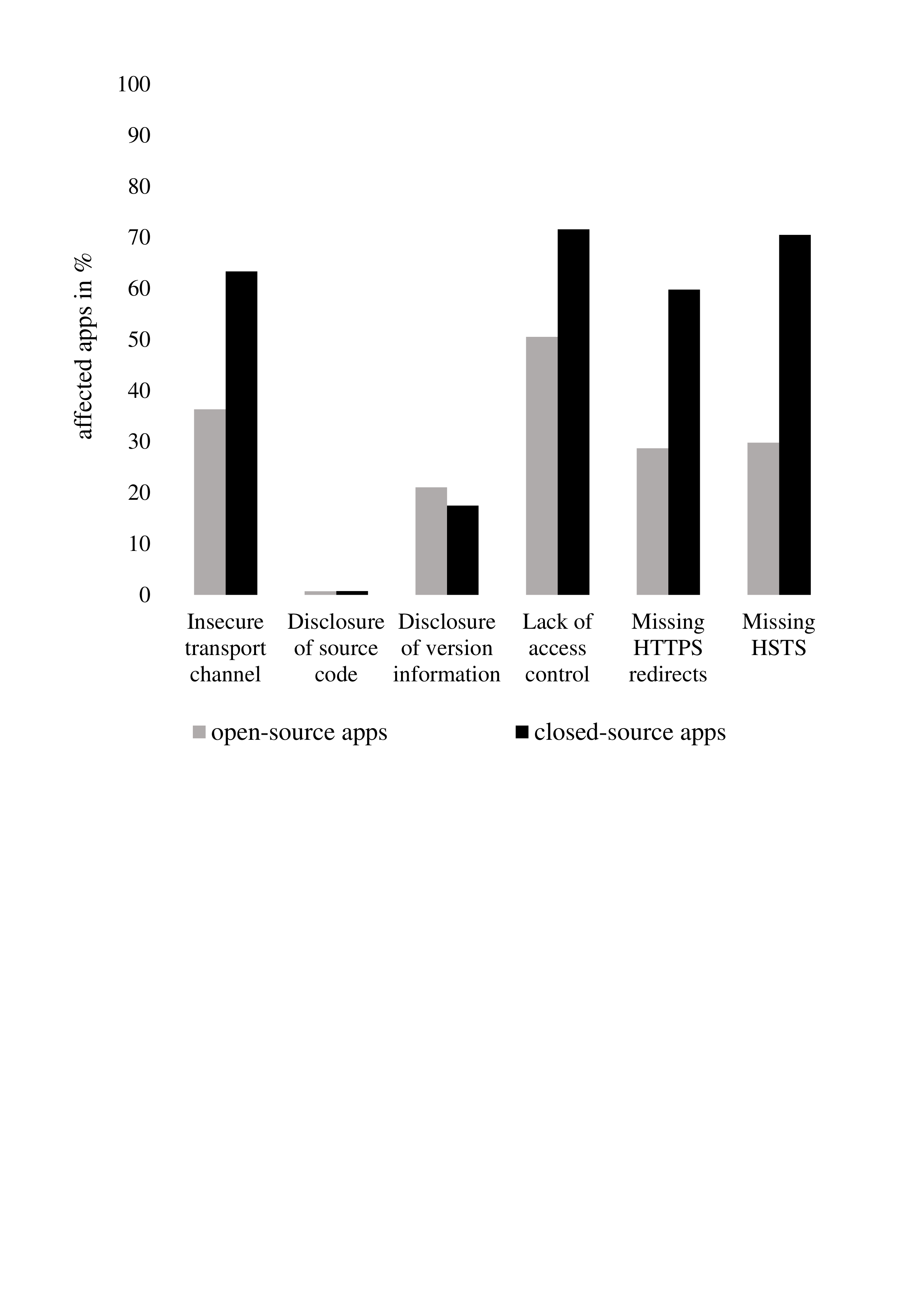}
\caption{Prevalence of app server smells in apps considering non-JSON communication}
\label{fig:p_smells}
\end{figure}

\subsection{Prevalence of Security Smells}
\label{sec:prevalence-security-smells}
This subsection answers \textbf{RQ$_{1}$}:
\emph{What is the prevalence of the server side security smells in web communication?}
In~\autoref{fig:p_smells_json} and~\autoref{fig:p_smells} we report  on the relative prevalence of app server security smells in apps for JSON and non-JSON web services, respectively.
In~\autoref{fig:p_smells_json}, the vertical axis indicates the percentage of apps that suffer from a specific app server security smell.
In the following, we discuss the findings from different perspectives, \ie security smell categories, software development model, and technology.

\subsubsection{By Security Smell Category}
We report findings and provide actionable advice to mitigate the issue for each security smell.

\emph{Insecure transport channel.}
Communication through an insecure transport channel is prone to data leaks and manipulation, \eg an adversary could alter conversations.
Hence, practitioners should avoid HTTP and instead focus on the secure HTTPS.
Third-party libraries that require HTTP should be replaced with ones that support secure communication.
With respect to URLs from open-source apps, we found that 582 non-JSON app servers (50\%) did not use protected communication.
Fortunately, this is not the case for JSON app servers:
only six JSON app servers (10\%) used plain text communication.
We found worse results in closed-source communication.
Secure communication was usually unavailable, \ie 5\,639 non-JSON app servers (71\%) used HTTP.
A total of 245 JSON app servers were not protected (50\%).

\begin{figure}
\centering
\includegraphics[scale = 0.39, trim = 2cm 15.5cm 2cm 1.25cm]{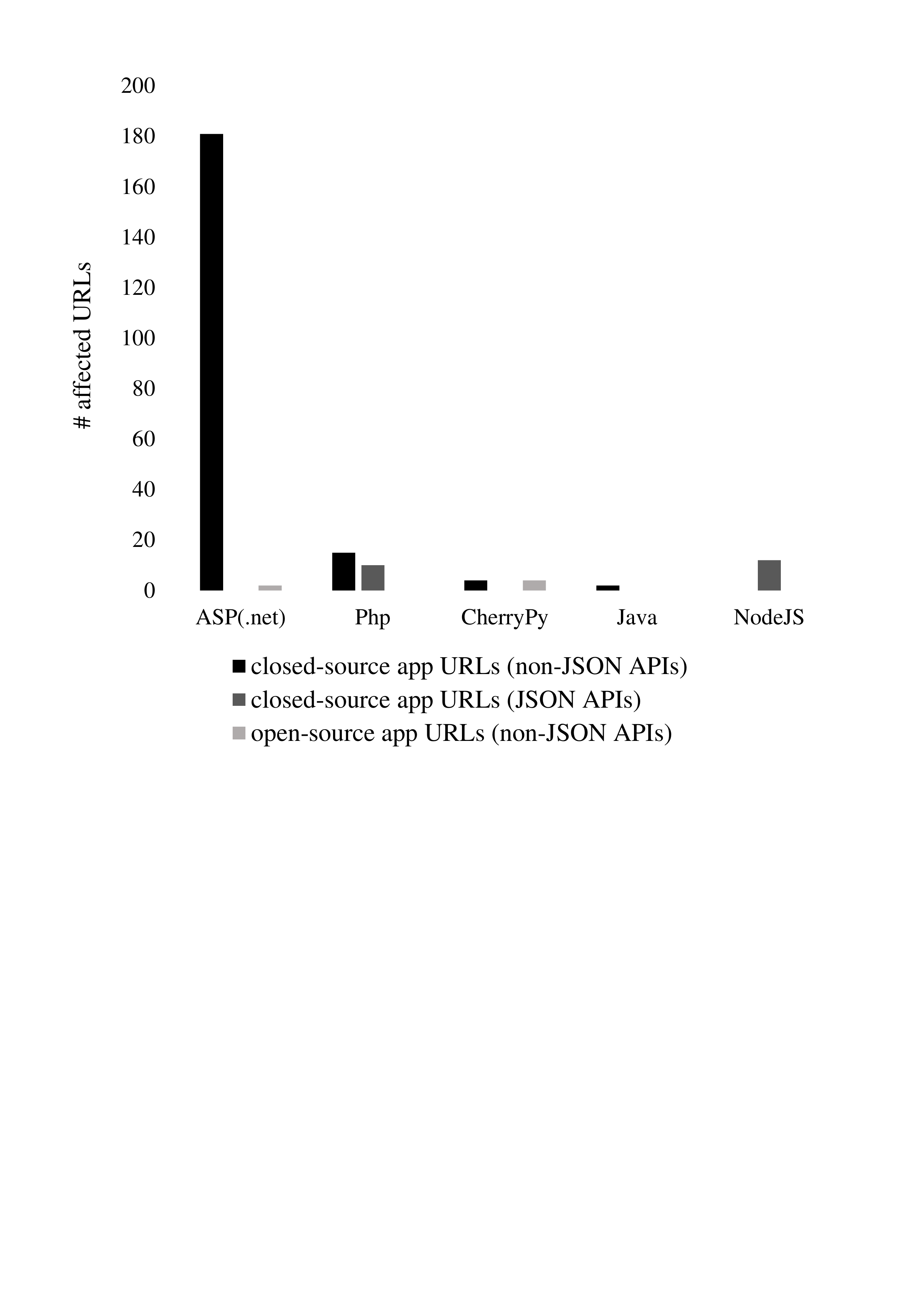}
\caption{Frameworks that caused code leaks}
\label{fig:p_s01}
\end{figure}

\emph{Disclosure of source code.}
Leaked code is valuable for adversaries to plot their attacks, or for competitors to glimpse into the source code and the architecture.
Therefore, administrators should disable verbose error messages on production environments and review the default settings.
We could identify stack traces from five different server frameworks, \ie ASP(.net), CherryPy, Java, NodeJS, and Php.
As we can see in~\autoref{fig:p_s01}, URLs from closed-source applications suffer the most from code leaks, \ie we found 225 instances (2.7\%) where 182 instances can be assigned to the ASP(.net) framework.
Considering URLs used in open-source software, we only found six instances (0.5\%) primarily caused by ASP(.net) and CherryPy.

\begin{figure}
\centering
\includegraphics[scale = 0.39, trim = 2cm 15.5cm 2cm 1.25cm]{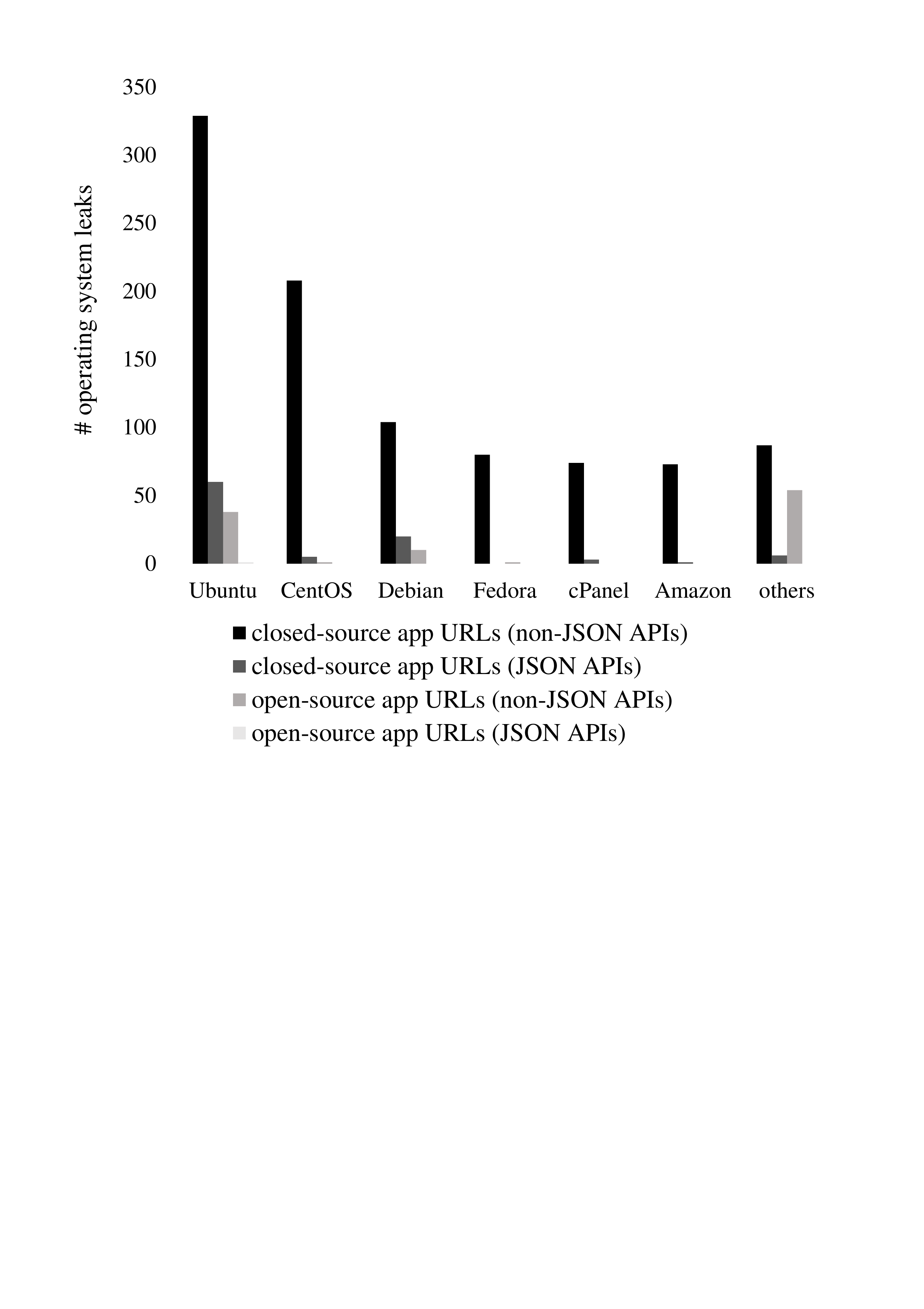}
\caption{Disclosure of operating system information}
\label{fig:p_s02c}
\end{figure}

\emph{Disclosure of version information.}
The knowledge of what exact software runs on a server is crucial for successful attacks.
Consequently, administrators should disable the self-promotion of services and review their default settings.
In~\autoref{fig:p_s02c}, we present the found operating system leaks in app servers, where the y-axis denotes the number of leaks we found.
We found 1\,155 operating system leaks in our dataset.
Ubuntu and Debian are the most prevalent operating systems for JSON app servers, and CentOS is rather used for non-JSON app servers.
Customized Linux distributions, \ie cPanel and Amazon, are less commonly used among web application developers.
\begin{figure}
\centering
\includegraphics[scale = 0.39, trim = 2cm 15.5cm 2cm 1.25cm]{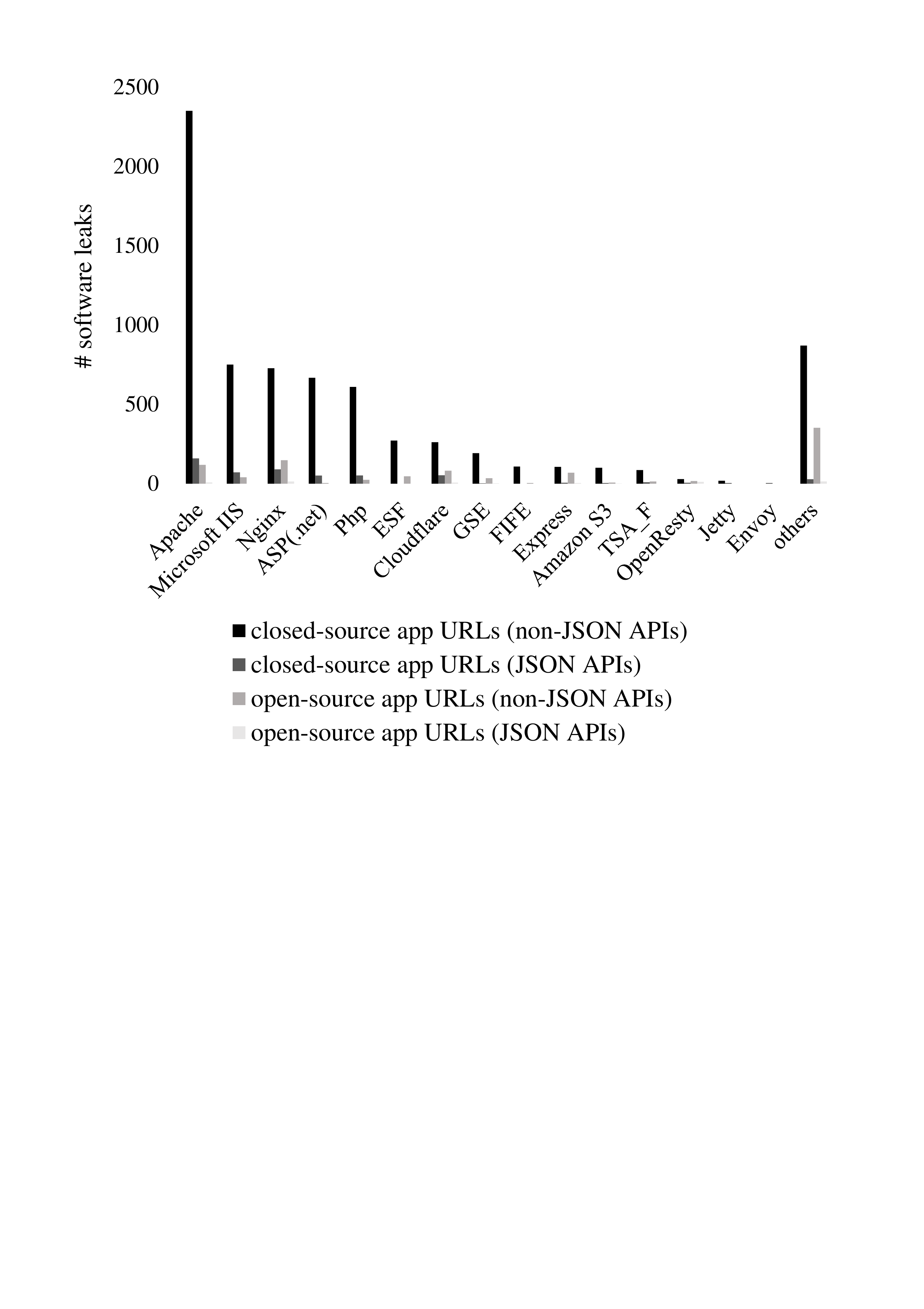}
\caption{Disclosure of service information}
\label{fig:p_s02b}
\end{figure}
In~\autoref{fig:p_s02b}, we present the found service leaks in app servers, where the y-axis denotes the number of leaks we found.
We found 8\,707 service leaks in our dataset, including servers that pack up to three leaks into a single HTTP response.
Open-source and closed-source software behave similarly, \ie Apache and Nginx are among the top three web application gateway servers used, but Microsoft services, \ie Microsoft IIS and ASP(.net), remain a preferred choice for closed-source developers.
Interestingly, the web security provider Cloudflare is used not only for numerous closed-source apps, but also for open-source apps, as we expect, due to their free plans.
Furthermore, the service leaks indicate that most of the app servers do not use the Google Cloud API (ESF) or storage services such as Amazon S3.
\begin{figure}
\centering
\includegraphics[scale = 0.39, trim = 2cm 15.5cm 2cm 1.25cm]{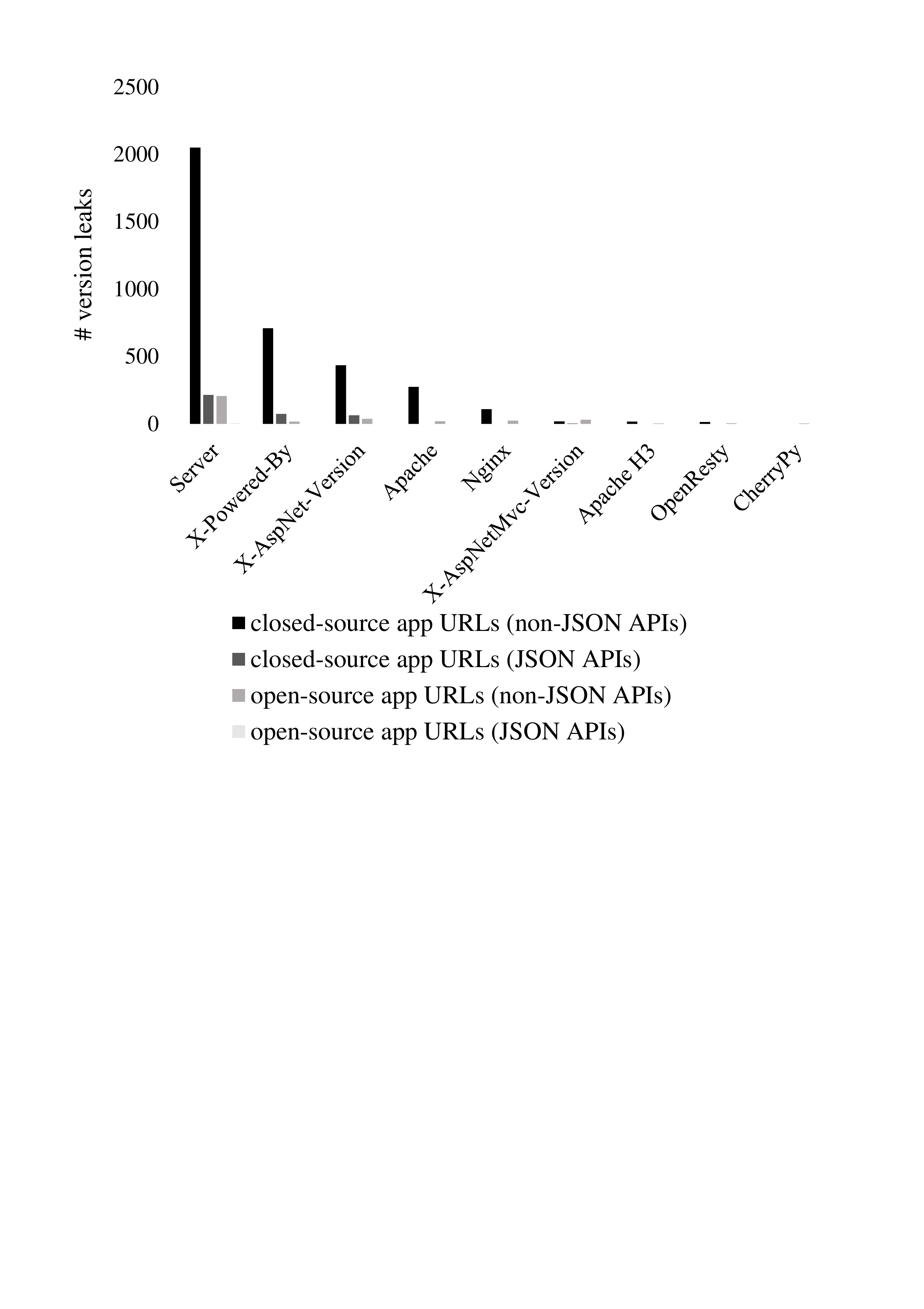}
\caption{Disclosure of version information}
\label{fig:p_s02}
\end{figure}
In~\autoref{fig:p_s02}, we present the found version leaks in app servers, where the y-axis denotes the number of leaks we found.
We found 3\,992 closed-source and 359 open-source software leaks in our dataset.
Most version leaks occur for both closed-source and open-source app servers in the HTTP header field \code{Server}, followed by \code{X-Powered-By}, and \code{X-AspNet-Version}.
The leaks in HTTP bodies, \ie \code{Apache}, \code{Nginx}, \code{Apache H3}, \code{OpenResty}, and \code{CherryPy} are less prevalent than those found in the headers.

\emph{Lack of access control.}
Unprotected information can be accessed by everyone on the internet.
Since apps usually provide experiences tailored to each user, their servers should use well known authentication schemes to prevent leaks of personal data.
We encountered 53 HTTP authentication errors for closed-source non-JSON app servers, and 28 errors for open-source non-JSON app servers.
We did not find any such errors for open-source or closed-source JSON app servers.
However, there exist JSON web applications that returned arbitrary authorization errors in the JSON format, \eg using OAuth instead of the HTTP mechanism.

\emph{Missing HTTPS redirects.}
\begin{figure}
\centering
\includegraphics[scale = 0.39, trim = 2cm 15.5cm 2cm 1.25cm]{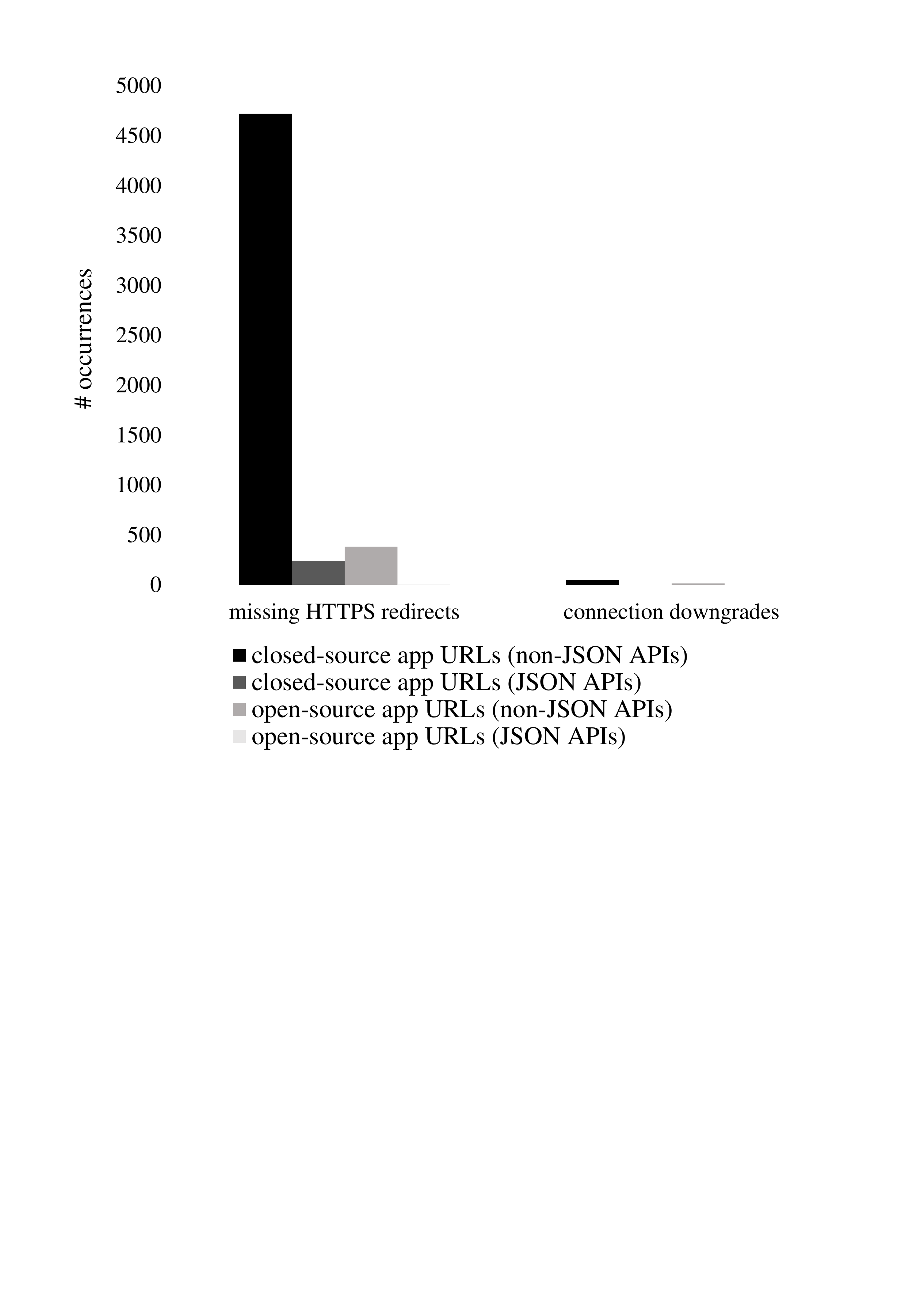}
\caption{Missing HTTPS redirects in app servers}
\label{fig:p_s04}
\end{figure}
Missing redirects leave flawed or outdated clients vulnerable to eavesdropping.
Redirects should always be set in place, if a server has ever been accessible through the insecure HTTP protocol.
Redirects can be chained, but they should be used sparingly.
As shown in~\autoref{fig:p_s04}, we found server responses with missing HTTPS redirects in the URLs from 4\,961 closed-source apps and from 387 open-source apps.
Fortunately, we did not find any HTTPS to HTTP connection downgrades in JSON app servers, but we found 48 for closed-source non-JSON app servers and 15 in open-source non-JSON app servers.
Concerning forwarded requests, closed-source app servers forwarded the requests on average 1.3 times, open-source non-JSON app servers 1.5 times, and open-source JSON app servers once.
We found two request loops, \ie infinite redirects from a destination to itself, in each open-source and closed-source app servers.
Without the request loops, open-source app servers redirected a request up to three times, and closed-source app servers up to seven times.

\emph{Missing HSTS.}
\begin{figure}
\centering
\includegraphics[scale = 0.39, trim = 2cm 15.5cm 2cm 1.25cm]{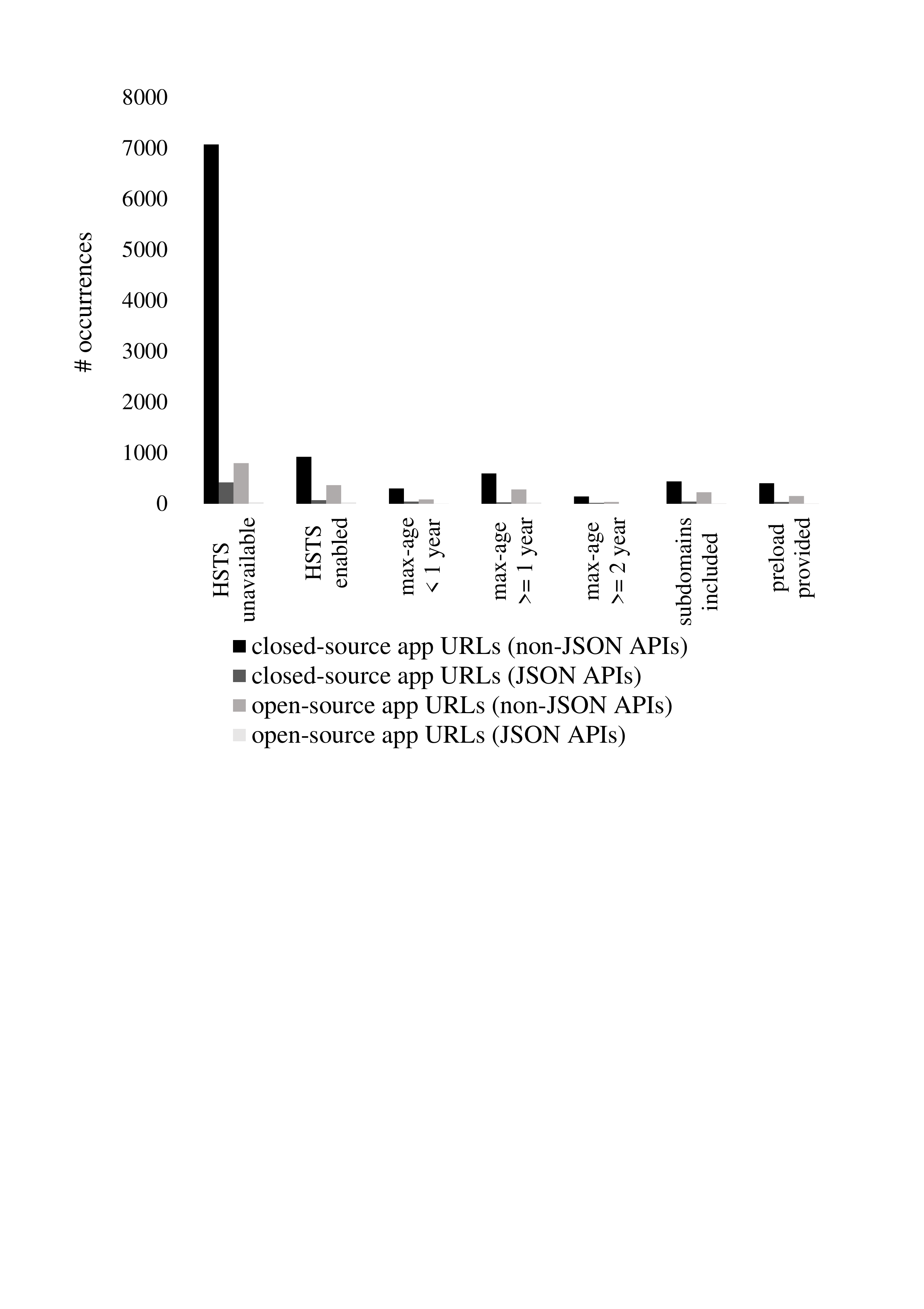}
\caption{Missing HSTS protection for app servers}
\label{fig:p_s05}
\end{figure}
App servers without proper support for HSTS expose users to eavesdropping due to possible HTTPS to HTTP connection downgrades.
Therefore, servers should deploy this feature to every subdomain and request the client side caching of this setting for at least one year.
Ultimately, the protected URLs should be added to the publicly available HSTS preload list that is included in all major browsers.
As shown in~\autoref{fig:p_s05}, we found 7\,494 closed-source app servers and 833 open-source app servers that miss HSTS HTTP headers.
Only a minority of the connections are protected, that is 397 (34\%) of all open-source app servers and 992 (12\%) of all closed-source app servers.
Contrary to recommended practices,\footnote{Google Chrome HSTS preload list submission form, \url{https://hstspreload.org/}} 432 app servers use \code{max-age} values shorter than one year, and 785 do not use the preload feature.
In other words, 31\% of the app servers that support HSTS have not sufficiently configured the protection for subdomains, and 57\% lack the preload feature that enforces security already for the first request.

\subsubsection{By Software Development Model}
We report findings for two different software development models, \ie the open-source and the closed-source software development model.
We can clearly see in~\autoref{fig:p_smells_json} and~\autoref{fig:p_smells} that closed-source apps generally suffer from more security smells than open-source apps.
Especially \smellText{Lack of access control} and \smellText{Missing HSTS} appear in the communication of almost all closed-source apps.
Moreover, the three smells \smellText{Insecure transport channel}, \smellText{Missing HTTPS redirects} and \smellText{Disclosure of version information} are less frequent, but exist still in more than 52\% of all closed-source apps and in more than 39\% of all open-source apps.
Interestingly, \smellText{Disclosure of source code} only emerges in closed-source app communication.

\subsubsection{By Technology}
We report our findings for two different technologies, \ie the JSON and non-JSON-based web communication.
According to~\autoref{fig:p_smells_json}, access control and unprotected HTTP communication constitute major threats for apps that use JSON web services.
However, apps that do not rely on JSON communication are apparently more robust against security smells:
such apps are on average about 19\% less affected by them.
Code leaks primarily occurred in JSON communication.
For instance, \smellText{Disclosure of source code} only exists in less than 1\% of the apps that use non-JSON web services, whereas it is more than 8\% for the apps that use regular JSON web services.
We only found code leaks in JSON app servers
that use the Php or NodeJS framework, but in contrast, we found code leaks in non-JSON servers from almost every major framework.

\subsubsection{Summary}
App server security smells pose a severe threat.
Most security smells exist in more than 25\% of all apps, regardless whether the app is open-source or closed-source, and whether it uses a JSON or non-JSON app server.
Particularly alarming is the finding that apps using JSON app servers suffer 1.5 times more from app server security smells than non-JSON apps, and even worse, closed-source applications suffer 1.6 times more compared to open-source applications.

More than 50\% of the servers accessed by mobile apps use unprotected HTTP communication.
Since smart devices are becoming rather personal assistants, they carry much sensitive information that needs adequate protection.

Misconfigured app servers cause code leaks.
Although only little code is revealed at a time, an attacker can replay requests and alter parameters to reconstruct the architecture and logic behind the service.
Such information eases the search for bugs in the code.

The leaked information is devastating. 
Although intended for publicity purposes, the currently leaked data reveals very often not only the operating system running on the server, but also the installed services and their version number.
Such information can be entered into vulnerability databases to find suitable security issues that could be exploited.

Based on our results, access control for JSON app servers is currently not implemented with HTTP status codes, but instead with arbitrary replies.
A standardized approach would help in creating more service independent apps, and at the same time default authorization templates could be used from back-end developers.

HTTPS redirects are usually inexistent for HTTP-based app servers.
Even worse, some downgrade a HTTPS connection to an insecure HTTP connection.
Moreover, redirect loops exist occasionally, and few redirect implementations use more than five redirects which is not recommended by RFC2068.\footnote{\url{RFC2068, HTTP/1.1, https://tools.ietf.org/html/rfc2068\#section-10.3}}

Finally, HSTS is only set up for a minority of app servers, and for those it is common to have weak configurations.

In conclusion, we see that security smells are very prevalent in app servers.
In fact, every app references on average more than three servers that suffer from at least one of these smells.

\subsection{Maintenance of Server Infrastructure}
\label{sec:maintenance-of-server-infrastructure}
In order to answer \textbf{RQ$_{2}$}:
\rqsix, we investigate maintenance operations performed on the servers used by mobile apps.
In particular, we are interested whether app server administrators have updated their infrastructure within the time period of 14 months, and if we see a correlation between the number of identified security smells and the quality of server maintenance.
The selected duration of more than a year covers multiple bug fixes including major releases of common server software, \eg Apache, Microsoft IIS, or PHP.
We accessed the URLs by sending an HTTP GET request, and stored their HTTP header responses twice, \ie once in June-2019 and once in August-2020.
We can only compare version numbers between the two datasets if we received some version information in the HTTP \code{Server} header.
As a result, the data in this section are based on fewer responses, \ie from 309 open-source (JSON and non-JSON) app server URLs (25\%) and 3\,006 closed-source (JSON and non-JSON) app server URLs (35\%).

During our manual analysis of the first 100 entries, we encountered eight different scenarios:
\begin{inparaenum}[i)]
\item no updates have been applied, \ie the software name and version remains identical, 
\item the version has been downgraded, \ie the software name remains identical, but the version number decreased,
\item the version has been upgraded, \ie the software name remains, but the version number increased,
\item the version leak has been closed, \ie the software name remains, but the version number is not anymore available,
\item the environment has changed, \ie the software has been replaced and it might use a different versioning scheme,
\item Cloudflare protection has been enabled, \ie the server has moved behind a Cloudflare protection gateway and does not anymore leak version information,
\item server spawned, \ie we received no software name in the first run, but we received one in the second run,
\item server shutdown, \ie we received a software name in the first run, but not anymore in the second run.
\end{inparaenum}
We could not gather security-related changes for 1\,254 app server URLs for several reasons:
\begin{inparaenum}[i)]
\item new server instances have been spawned without prior knowledge of software configurations, 
\item existing server instances have been shutdown without the possibility to find any changes, or 
\item the environment has changed using a different versioning scheme.
\end{inparaenum}

\begin{figure}
\centering
\includegraphics[scale = 0.39, trim = 2cm 15.5cm 2cm 1.25cm]{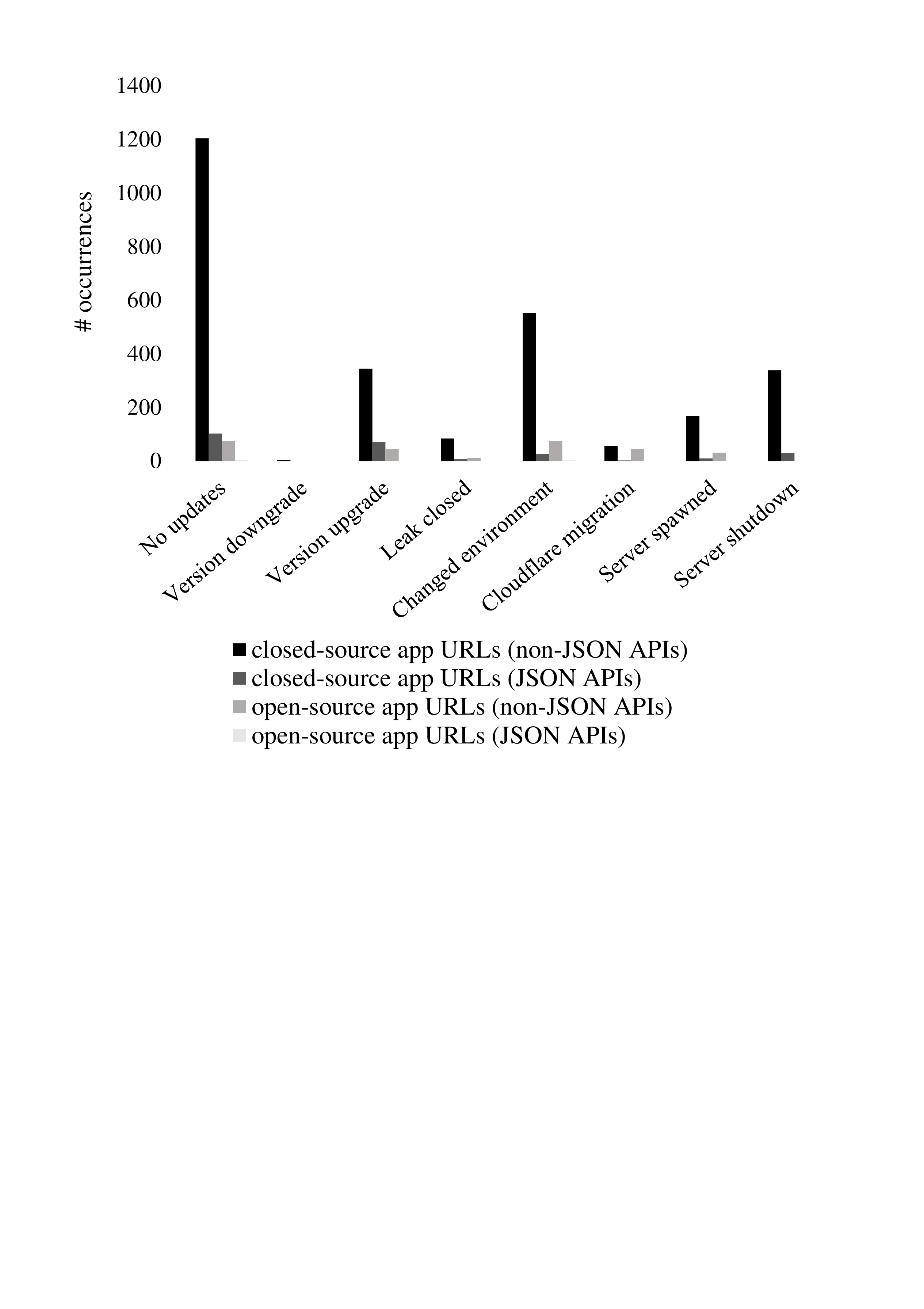}
\caption{Configuration changes of app servers after 14 months}
\label{fig:p_u01}
\end{figure}

\subsubsection{Configuration Changes}
In~\autoref{fig:p_u01}, we show the results.
From the app servers that leaked versioning information, by far most closed-source non-JSON app servers did not undergo any changes to the server software.
Closed-source JSON app server infrastructure seems to be updated more frequently, however the majority still do not provide any updates.
The same is true for open-source software although less evident.
Version downgrades occurred sparsely, \ie four times, and not for JSON app servers.
Only a fraction of the leaking servers, \ie 103 (4\%), has been configured to mitigate the leaks.
Interestingly, environment changes occur more frequently for open-source non-JSON app servers than no updates at all.
In other words, open-source developers seem to replace app servers rather then updating them.
Moreover, Cloudflare support has been enabled for 104 app servers, \ie for 45 open-source URLs and for 59 closed-source URLs.
Finally, more servers are shut down than spawned.

\begin{figure}
\centering
\includegraphics[scale = 0.39, trim = 2cm 15.5cm 2cm 1.25cm]{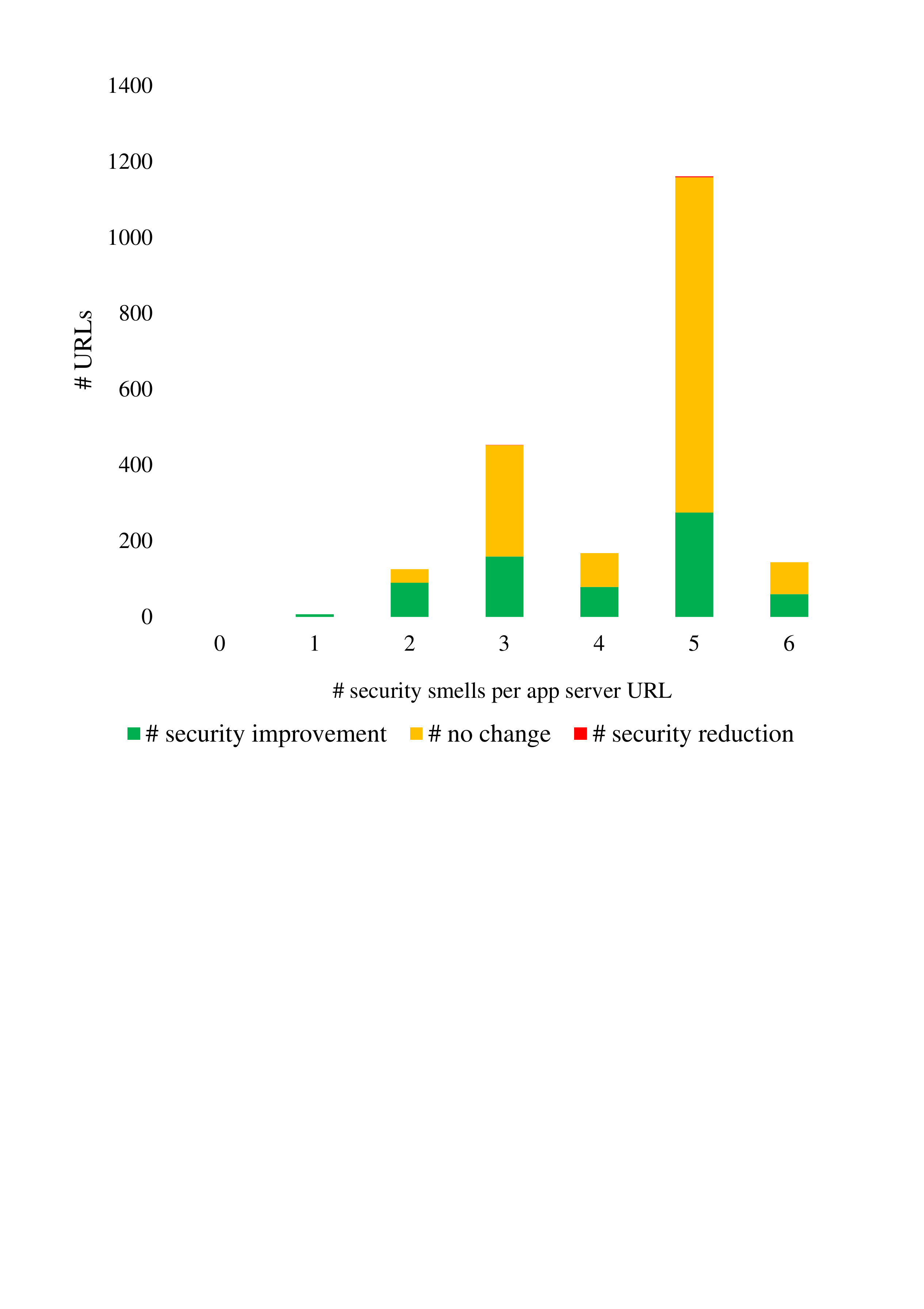}
\caption{Correlation between app server security smells and configuration changes}
\label{fig:p_oURL_cURL}
\end{figure}

\subsubsection{Correlation of Security Smells}
\autoref{fig:p_oURL_cURL} shows the correlation between app server security smells and administrative configuration changes.
For this figure, we consolidated all app server categories, \ie open-source, closed-source, JSON, non-JSON due to the limited number of elements in some of them.
The x-axis denotes the number of security smells from which a particular app server suffers, and the y-axis indicates how many such app servers exist in each category.
Based on the versioning information from 2\,061 URLs, we can see that app servers suffering from three or more smells are usually not well maintained, \ie they are set up once and then left alone.
Although security improvements, \ie version upgrades, the removal of versioning information, and the migration to Cloudflare appear more frequently in instances that suffer from more than one smell, they only affect a minority.
Security downgrades, \ie the change to a more dated version, appear only in app servers that massively suffer from security smells, \ie from three or more smells.

\subsubsection{Summary}
According to our findings, app servers are usually set up once and never touched again.
This paradigm introduces severe security risks due to outdated software running on publicly accessible interfaces.
Hence, sensitive user data could be exfiltrated when adversaries apply suitable exploits to such systems.
Luckily, version upgrades are much more common than version downgrades, although they cannot at all compensate for the lack of change.
We expect that downgrades were performed to circumvent new bugs or compatibility issues, because all downgrades considered only minor release changes, \eg from \emph{nginx} release 1.14.1 to 1.12.1.
Some developers shift to Cloudflare to protect their infrastructure especially for non-JSON app servers.

We conclude that app server security smells seem to be a good indicator for poor server maintenance.
In fact, the more smells an app server has the more likely it is that server maintenance processes are broken.

\section{Threats to Validity}
\label{sec:threats-to-validity}
\emph{Completeness.}
A major threat to validity is the completeness of the used dataset built from Android apps.
Although state of the art decompilation tools have been used, only about 37\% of all closed-source Android apps could be successfully decompiled for the subsequent analysis.
Of these decompiled apps, the analysis for 22\% could not finish in time and might have led to incomplete results.
Moreover, the analysis tool skipped the evaluation of bundled build scripts and XML resources that could have pointers to additional app servers.
This threat cannot be mitigated entirely, however the rather large and diverse set of included apps ensures that the results can be generalized.

\emph{Accuracy.}
Another important threat represents the accuracy of the used dataset.
According to the authors, the tool that has been used to build the dataset achieves a precision of 46\% and a recall of 80\%.
However, this performance is the result of a manual analysis of decompiled code performed by the authors which included only ten open-source and ten closed-source apps that comprised 22 web API URLs.
In particular, it reported several URLs unrelated to web APIs but to static HTML pages, and the tool occasionally reconstructed invalid requests.
In this work, we do not depend on accurate requests, \ie the investigated response headers are identical even for malformed requests.
In fact, most of the reconstructed requests contained placeholders that we could leverage to see whether the app servers leak sensitive information in case of errors.

\emph{Data collection.}
The collected data might contain duplicates or suffer from temporal issues.
Some requests we generated from the URL might have reached identical servers which ultimately lead to duplicated connection information in the result set.
Another problem is that of server side outages or configuration changes that temporarily cause unexpected or erroneous results.
To mitigate these threats, we filtered the URL list for duplicates, and we used rather long timeouts and a high retry count when we accessed the servers.

\emph{Selection bias.}
The data used for the investigation of server maintenance represents only a subset of the original dataset.
This is an immediate result of the many servers that do not leak any data.
Even more, for the qualitative analysis we require two responses, each containing versioning information.
In order to reduce the impact of these threats, we manually reviewed the first 100 server responses to ensure that we do not miss any version information.
We then designed the value extraction process for the individual version numbers based on the results of this initial exploration.

\emph{Recency.}
The data set contains apps that have been downloaded in 2018, and the corresponding metadata has been collected in 2019.
This might change the results due to improved development processes and tools.
However, recent works still identified a lack of security in web communication~\cite{Alashwali:2020,Hu:2021}.

\emph{Security risks.}
The risks associated with the security smells are not necessarily severe.
We do not know what and how much data the web services hoard, and many of the risks directly correlate with the confidentiality of the data.
Since we cannot easily obtain this information, we follow a defensive strategy, \ie we assume that every server might host at least some sensitive data.

\emph{Construct validity.}
There is a threat to construct validity through potential bias in our expectancy.

\section{Related Work}
\label{sec:related-work}
Related work primarily pertains to app analyses that have been  summarized by the concept of security code smells, data transmissions with a particular interest in web communication, and public service audits that improve the app server security.
We present relevant literature in each of these three research areas in the remainder of this section.

\subsection{Security Code Smells}
The research about security code smells investigates the metamorphosis from unfavorable code that could become a security threat.
Ghafari~\etal collected 46\,000 closed-source apps from the official Android market and investigated the nature and prevalence of common mistakes developers suffered.
For that purpose, they introduced the notion of a \emph{security code smell} and used it to identify 28 different security smells in five different categories~\cite{Ghafari:2017}.
They found that \emph{XSS-like Code Injection}, \emph{Dynamic Code Loading}, and \emph{Custom Scheme Channel} are the most prevalent smells, many of them leveraging inter-component communication features of the Android operating system.
As a result, Gadient~\etal started to study the prevalence of \emph{Inter-Component Communication (ICC)}-related security smells in more than 700 open-source apps, and found that security code smells that involve web communication prevail against others, and that such issues are often introduced with new feature updates of apps~\cite{Gadient:2018}.
Moreover, the manual investigation of 100 apps demonstrated the usefulness of their tool, \ie about 43\% of the reported smells were in fact vulnerabilities.
Since many of the newly discovered smells relied on responses from web applications, they consequently began to investigate the web API communication of mobile apps~\cite{Gadient:2020}.
The preliminary results of their static analysis tool, which has been used to mine security code smells were devastating:
In 3\,376 apps, they encountered credential leaks, excessive use of embedded languages such as SQL and JavaScript, insecure web communication including source-code and version information leaks to name a few.
As a matter of fact, they found that unprotected web communication is seven times more prevalent in closed-source apps compared to open-source apps, and that embedded code is used in web communication in more than 500 different apps.
Our work continues this research, \ie we investigate the server side prevalence of the reported security smells.

\subsection{Web Communication}
Web communication in apps is usually initiated by the client, \ie the app that sends a request to a specific server.
Therefore, apps can reveal interesting features used to establish such a connection.
For example, Zuo~\etal analyzed 5\,000 top-ranked apps in Google Play and identified 297\,780 URLs that they fed to the VirusTotal URL screening service~\cite{Zuo:2017}.
The service identified 8\,634 harmful URLs of which the majority related to malware (43\%), followed by malicious sites (37\%), and phishing (23\%).
Mendoza~\etal investigated the input validation constraints imposed by apps on outgoing requests to web API services from 10\,000 popular free apps from the Google Play Store of which 46\% suffered from inconsistencies that could be exploited by attackers~\cite{Mendoza:2018a}.
Such inconsistencies allowed them to access app-related databases through various injection attacks, \eg they could misuse an app's email address field for an SQL injection attack, because its value did not receive additional server side validation.
We found many similarities in the results of our work:
advertisement services were omnipresent and proper authentication measures were barely implemented.
For instance, access to personal information was protected by the sole use of a single attribute, \eg an email address or hardware-based identifier.

\subsection{App Server Security}
Finally, app server security focuses on server side problems, configuration or implementation.
Zuo~\etal found that 15\,098 app servers are subject to data leakage attacks~\cite{Zuo:2019}.
In particular, they suffer either from a broken key management, \ie the developers became confused about root and app keys, or from a broken permission configuration, \ie developers were overwhelmed when they had to choose appropriate permissions for their data.
They assume that this is a direct consequence of the utterly complex interfaces to configure such services designed for developers.
That is, Google even provides a language for developers to specify the desired user permissions.
With respect to web servers, Lavrenovs~\etal worked through responses of the top one million Alexa websites, and collected security-related information such as HSTS support, protection against cross site scripting, and other HTTP headers that might impose a security risk~\cite{Lavrenovs:2018}.
They found that website popularity is the major driver for security measures.
In fact, the implementation rates compared against the Alexa ranking reveal an exponential decline pattern, \ie all of their analysed security headers started to be much more prevalent in the top 50\,000 websites, and that ratio steeply increased towards the top websites.
Moreover, Mendoza~\etal found discrepancies between the use of such features in the mobile and desktop version of websites that enable various injection and spoofing attacks, although the affected websites remain in the realm of a few percent~\cite{Mendoza:2018b}.
Although we can confirm these results, according to our study a lack of security is much more prevalent in apps that use JSON app servers, especially in closed-source apps that are 11\% more susceptible to such issues than their open-source counterparts.

\section{Conclusion}
\label{sec:conclusion}
We analyzed the prevalence of six security smells in app servers and investigated the consequence of these smells from a security perspective.
We used an existing dataset that includes 9\,714 distinct URLs that were used in 3\,376 Android mobile apps.
We exercised the URLs twice over 14 months, and stored the HTTP headers and bodies.
We realized that the top three smells exist in more than 69\% of all tested apps, and that unprotected communication and server misconfigurations are very common.
Particularly alarming is the finding that apps using JSON app servers suffer 1.5 times more from app server security smells than non-JSON apps, and even worse, closed-source applications suffer 1.6 times more compared to open-source applications.
Moreover, source-code and version leaks, or the lack of update policies foster future attacks against these data centric systems.
We found that app server security smells are omnipresent and they indicate poor app server maintenance.

\begin{acks}
\blind{We gratefully acknowledge the financial support of the Swiss National Science Foundation for the project
``Agile Software Assistance'' (SNSF project No.\ 200020-181973, Feb.\ 1, 2019 - April 30, 2022).}
\end{acks}

\bibliographystyle{ACM-Reference-Format}
\bibliography{webapi}


\begin{thebibliography}{17}


\ifx \showCODEN    \undefined \def \showCODEN     #1{\unskip}     \fi
\ifx \showDOI      \undefined \def \showDOI       #1{#1}\fi
\ifx \showISBNx    \undefined \def \showISBNx     #1{\unskip}     \fi
\ifx \showISBNxiii \undefined \def \showISBNxiii  #1{\unskip}     \fi
\ifx \showISSN     \undefined \def \showISSN      #1{\unskip}     \fi
\ifx \showLCCN     \undefined \def \showLCCN      #1{\unskip}     \fi
\ifx \shownote     \undefined \def \shownote      #1{#1}          \fi
\ifx \showarticletitle \undefined \def \showarticletitle #1{#1}   \fi
\ifx \showURL      \undefined \def \showURL       {\relax}        \fi
\providecommand\bibfield[2]{#2}
\providecommand\bibinfo[2]{#2}
\providecommand\natexlab[1]{#1}
\providecommand\showeprint[2][]{arXiv:#2}

\bibitem[\protect\citeauthoryear{Alashwali, Szalachowski, and Martin}{Alashwali
  et~al\mbox{.}}{2020}]%
        {Alashwali:2020}
\bibfield{author}{\bibinfo{person}{Eman~Salem Alashwali},
  \bibinfo{person}{Pawel Szalachowski}, {and} \bibinfo{person}{Andrew Martin}.}
  \bibinfo{year}{2020}\natexlab{}.
\newblock \showarticletitle{Exploring {HTTPS} security inconsistencies: A
  cross-regional perspective}.
\newblock \bibinfo{journal}{\emph{Computers \& Security}}  \bibinfo{volume}{97}
  (\bibinfo{year}{2020}), \bibinfo{pages}{101975}.
\newblock


\bibitem[\protect\citeauthoryear{Gadient, Ghafari, Frischknecht, and
  Nierstrasz}{Gadient et~al\mbox{.}}{2018}]%
        {Gadient:2018}
\bibfield{author}{\bibinfo{person}{Pascal Gadient}, \bibinfo{person}{Mohammad
  Ghafari}, \bibinfo{person}{Patrick Frischknecht}, {and}
  \bibinfo{person}{Oscar Nierstrasz}.} \bibinfo{year}{2018}\natexlab{}.
\newblock \showarticletitle{Security Code Smells in {Android} {ICC}}.
\newblock \bibinfo{journal}{\emph{Empirical Software Engineering Special
  Issue}} (\bibinfo{year}{2018}).
\newblock
\urldef\tempurl%
\url{https://doi.org/10.1007/s10664-018-9673-y}
\showDOI{\tempurl}


\bibitem[\protect\citeauthoryear{Gadient, Ghafari, Tarnutzer, and
  Nierstrasz}{Gadient et~al\mbox{.}}{2020}]%
        {Gadient:2020}
\bibfield{author}{\bibinfo{person}{Pascal Gadient}, \bibinfo{person}{Mohammad
  Ghafari}, \bibinfo{person}{Marc-Andrea Tarnutzer}, {and}
  \bibinfo{person}{Oscar Nierstrasz}.} \bibinfo{year}{2020}\natexlab{}.
\newblock \showarticletitle{Web {APIs} in {Android} through the Lens of
  Security}. In \bibinfo{booktitle}{\emph{2020 IEEE 27th International
  Conference on Software Analysis, Evolution and Reengineering (SANER)}}. IEEE,
  \bibinfo{pages}{13--22}.
\newblock


\bibitem[\protect\citeauthoryear{Ghafari, Gadient, and Nierstrasz}{Ghafari
  et~al\mbox{.}}{2017}]%
        {Ghafari:2017}
\bibfield{author}{\bibinfo{person}{M. Ghafari}, \bibinfo{person}{P. Gadient},
  {and} \bibinfo{person}{O. Nierstrasz}.} \bibinfo{year}{2017}\natexlab{}.
\newblock \showarticletitle{Security Smells in {Android}}. In
  \bibinfo{booktitle}{\emph{2017 IEEE 17th International Working Conference on
  Source Code Analysis and Manipulation (SCAM)}}. \bibinfo{pages}{121--130}.
\newblock
\urldef\tempurl%
\url{https://doi.org/10.1109/SCAM.2017.24}
\showDOI{\tempurl}


\bibitem[\protect\citeauthoryear{Gordon, Kim, Perkins, Gilham, Nguyen, and
  Rinard}{Gordon et~al\mbox{.}}{2015}]%
        {Gordon:2015}
\bibfield{author}{\bibinfo{person}{Michael~I Gordon}, \bibinfo{person}{Deokhwan
  Kim}, \bibinfo{person}{Jeff~H Perkins}, \bibinfo{person}{Limei Gilham},
  \bibinfo{person}{Nguyen Nguyen}, {and} \bibinfo{person}{Martin~C Rinard}.}
  \bibinfo{year}{2015}\natexlab{}.
\newblock \showarticletitle{Information flow analysis of {Android} applications
  in {DroidSafe}.}. In \bibinfo{booktitle}{\emph{NDSS}},
  Vol.~\bibinfo{volume}{15}. \bibinfo{pages}{110}.
\newblock


\bibitem[\protect\citeauthoryear{Hu, Asghar, and Brownlee}{Hu
  et~al\mbox{.}}{2021}]%
        {Hu:2021}
\bibfield{author}{\bibinfo{person}{Qinwen Hu}, \bibinfo{person}{Muhammad~Rizwan
  Asghar}, {and} \bibinfo{person}{Nevil Brownlee}.}
  \bibinfo{year}{2021}\natexlab{}.
\newblock \showarticletitle{A large-scale analysis of {HTTPS} deployments:
  Challenges, solutions, and recommendations}.
\newblock \bibinfo{journal}{\emph{Journal of Computer Security}}
  \bibinfo{number}{Preprint} (\bibinfo{year}{2021}), \bibinfo{pages}{1--26}.
\newblock


\bibitem[\protect\citeauthoryear{Lavrenovs and Melón}{Lavrenovs and
  Melón}{2018}]%
        {Lavrenovs:2018}
\bibfield{author}{\bibinfo{person}{Arturs Lavrenovs} {and}
  \bibinfo{person}{F.~Jesús~Rubio Melón}.} \bibinfo{year}{2018}\natexlab{}.
\newblock \showarticletitle{{HTTP} security headers analysis of top one million
  websites}. In \bibinfo{booktitle}{\emph{2018 10th International Conference on
  Cyber Conflict (CyCon)}}. \bibinfo{pages}{345--370}.
\newblock
\urldef\tempurl%
\url{https://doi.org/10.23919/CYCON.2018.8405025}
\showDOI{\tempurl}


\bibitem[\protect\citeauthoryear{Mendoza, Chinprutthiwong, and Gu}{Mendoza
  et~al\mbox{.}}{2018}]%
        {Mendoza:2018b}
\bibfield{author}{\bibinfo{person}{Abner Mendoza}, \bibinfo{person}{Phakpoom
  Chinprutthiwong}, {and} \bibinfo{person}{Guofei Gu}.}
  \bibinfo{year}{2018}\natexlab{}.
\newblock \showarticletitle{Uncovering {HTTP} Header Inconsistencies and the
  Impact on Desktop/Mobile Websites}. In \bibinfo{booktitle}{\emph{Proceedings
  of the 2018 World Wide Web Conference}} (Lyon, France)
  \emph{(\bibinfo{series}{WWW '18})}. \bibinfo{publisher}{International World
  Wide Web Conferences Steering Committee}, \bibinfo{address}{Republic and
  Canton of Geneva, CHE}, \bibinfo{pages}{247–256}.
\newblock
\showISBNx{9781450356398}
\urldef\tempurl%
\url{https://doi.org/10.1145/3178876.3186091}
\showDOI{\tempurl}


\bibitem[\protect\citeauthoryear{Mendoza and Gu}{Mendoza and Gu}{2018}]%
        {Mendoza:2018a}
\bibfield{author}{\bibinfo{person}{Abner Mendoza} {and} \bibinfo{person}{Guofei
  Gu}.} \bibinfo{year}{2018}\natexlab{}.
\newblock \showarticletitle{Mobile application web {API} reconnaissance:
  Web-to-mobile inconsistencies \& vulnerabilities}. In
  \bibinfo{booktitle}{\emph{2018 IEEE Symposium on Security and Privacy (SP)}}.
  IEEE, \bibinfo{pages}{756--769}.
\newblock


\bibitem[\protect\citeauthoryear{Possemato and Fratantonio}{Possemato and
  Fratantonio}{2020}]%
        {Possemato:2020}
\bibfield{author}{\bibinfo{person}{Andrea Possemato} {and}
  \bibinfo{person}{Yanick Fratantonio}.} \bibinfo{year}{2020}\natexlab{}.
\newblock \showarticletitle{Towards {HTTPS} Everywhere on {Android}: We Are Not
  There Yet}. In \bibinfo{booktitle}{\emph{29th {USENIX} Security Symposium
  ({USENIX} Security 20)}}. \bibinfo{pages}{343--360}.
\newblock


\bibitem[\protect\citeauthoryear{Rahman, Parnin, and Williams}{Rahman
  et~al\mbox{.}}{2019}]%
        {Rahman:2019}
\bibfield{author}{\bibinfo{person}{Akond Rahman}, \bibinfo{person}{Chris
  Parnin}, {and} \bibinfo{person}{Laurie Williams}.}
  \bibinfo{year}{2019}\natexlab{}.
\newblock \showarticletitle{The Seven Sins: Security Smells in Infrastructure
  As Code Scripts}. In \bibinfo{booktitle}{\emph{Proceedings of the 41st
  International Conference on Software Engineering}} (Montreal, Quebec, Canada)
  \emph{(\bibinfo{series}{ICSE '19})}. \bibinfo{publisher}{IEEE Press},
  \bibinfo{address}{Piscataway, NJ, USA}, \bibinfo{pages}{164--175}.
\newblock
\urldef\tempurl%
\url{https://doi.org/10.1109/ICSE.2019.00033}
\showDOI{\tempurl}


\bibitem[\protect\citeauthoryear{Rapoport, Suter, Wittern, Lh\'{o}tak, and
  Dolby}{Rapoport et~al\mbox{.}}{2017}]%
        {Rapoport:2017}
\bibfield{author}{\bibinfo{person}{Marianna Rapoport},
  \bibinfo{person}{Philippe Suter}, \bibinfo{person}{Erik Wittern},
  \bibinfo{person}{Ond\v{r}ej Lh\'{o}tak}, {and} \bibinfo{person}{Julian
  Dolby}.} \bibinfo{year}{2017}\natexlab{}.
\newblock \showarticletitle{Who You Gonna Call?: Analyzing Web Requests in
  {Android} Applications}. In \bibinfo{booktitle}{\emph{Proceedings of the 14th
  International Conference on Mining Software Repositories}} (Buenos Aires,
  Argentina) \emph{(\bibinfo{series}{MSR '17})}. \bibinfo{publisher}{IEEE
  Press}, \bibinfo{address}{Piscataway, NJ, USA}, \bibinfo{pages}{80--90}.
\newblock
\showISBNx{978-1-5386-1544-7}
\urldef\tempurl%
\url{https://doi.org/10.1109/MSR.2017.11}
\showDOI{\tempurl}


\bibitem[\protect\citeauthoryear{Tang, Ouyang, and Tsai}{Tang
  et~al\mbox{.}}{2015}]%
        {Tang:2015}
\bibfield{author}{\bibinfo{person}{Longji Tang}, \bibinfo{person}{Liubo
  Ouyang}, {and} \bibinfo{person}{Wei-Tek Tsai}.}
  \bibinfo{year}{2015}\natexlab{}.
\newblock \showarticletitle{Multi-factor web {API} security for securing Mobile
  Cloud}. In \bibinfo{booktitle}{\emph{2015 12th International Conference on
  Fuzzy Systems and Knowledge Discovery (FSKD)}}. \bibinfo{pages}{2163--2168}.
\newblock
\urldef\tempurl%
\url{https://doi.org/10.1109/FSKD.2015.7382287}
\showDOI{\tempurl}


\bibitem[\protect\citeauthoryear{Wittern, Ying, Zheng, Dolby, and
  Laredo}{Wittern et~al\mbox{.}}{2017}]%
        {Wittern:2017}
\bibfield{author}{\bibinfo{person}{Erik Wittern}, \bibinfo{person}{Annie~T.T.
  Ying}, \bibinfo{person}{Yunhui Zheng}, \bibinfo{person}{Julian Dolby}, {and}
  \bibinfo{person}{Jim~A. Laredo}.} \bibinfo{year}{2017}\natexlab{}.
\newblock \showarticletitle{Statically Checking Web {API} Requests in
  JavaScript}. In \bibinfo{booktitle}{\emph{2017 IEEE/ACM 39th International
  Conference on Software Engineering (ICSE)}}. \bibinfo{pages}{244--254}.
\newblock
\urldef\tempurl%
\url{https://doi.org/10.1109/ICSE.2017.30}
\showDOI{\tempurl}


\bibitem[\protect\citeauthoryear{Zhou, Wu, Wang, and Jiang}{Zhou
  et~al\mbox{.}}{2015}]%
        {Zhou:2015}
\bibfield{author}{\bibinfo{person}{Yajin Zhou}, \bibinfo{person}{Lei Wu},
  \bibinfo{person}{Zhi Wang}, {and} \bibinfo{person}{Xuxian Jiang}.}
  \bibinfo{year}{2015}\natexlab{}.
\newblock \showarticletitle{Harvesting developer credentials in {Android}
  apps}. In \bibinfo{booktitle}{\emph{WISEC}}. \bibinfo{pages}{1--12}.
\newblock


\bibitem[\protect\citeauthoryear{Zuo and Lin}{Zuo and Lin}{2017}]%
        {Zuo:2017}
\bibfield{author}{\bibinfo{person}{Chaoshun Zuo} {and}
  \bibinfo{person}{Zhiqiang Lin}.} \bibinfo{year}{2017}\natexlab{}.
\newblock \showarticletitle{SMARTGEN: Exposing Server {URL}s of Mobile Apps
  With Selective Symbolic Execution}. In \bibinfo{booktitle}{\emph{Proceedings
  of the 26th International Conference on World Wide Web}} (Perth, Australia)
  \emph{(\bibinfo{series}{WWW '17})}. \bibinfo{publisher}{International World
  Wide Web Conferences Steering Committee}, \bibinfo{address}{Republic and
  Canton of Geneva, Switzerland}, \bibinfo{pages}{867--876}.
\newblock
\showISBNx{978-1-4503-4913-0}
\urldef\tempurl%
\url{https://doi.org/10.1145/3038912.3052609}
\showDOI{\tempurl}


\bibitem[\protect\citeauthoryear{Zuo, Lin, and Zhang}{Zuo
  et~al\mbox{.}}{2019}]%
        {Zuo:2019}
\bibfield{author}{\bibinfo{person}{Chaoshun Zuo}, \bibinfo{person}{Zhiqiang
  Lin}, {and} \bibinfo{person}{Yinqian Zhang}.}
  \bibinfo{year}{2019}\natexlab{}.
\newblock \showarticletitle{Why Does Your Data Leak? Uncovering the Data
  Leakage in Cloud from Mobile Apps}. In \bibinfo{booktitle}{\emph{2019 IEEE
  Symposium on Security and Privacy (SP)}}. \bibinfo{pages}{1296--1310}.
\newblock
\urldef\tempurl%
\url{https://doi.org/10.1109/SP.2019.00009}
\showDOI{\tempurl}


\end{thebibliography}

\end{document}